\documentclass[fleqn,usenatbib]{mnras}

\usepackage{newtxtext,newtxmath}
\usepackage{footnote}
\usepackage{adjustbox}
\usepackage{lscape}
\usepackage{numprint}
\usepackage{graphicx}
\usepackage[T1]{fontenc}
\usepackage{ae,aecompl}

\usepackage{graphicx}	
\usepackage{amsmath}	
\usepackage{amssymb}	

\newcommand{\LSO}{La Silla Observatory}
\newcommand{\PAR}{Paranal Observatory}
\newcommand{\kepler}{{\it Kepler}}

\newcommand{\tess}{{\it TESS}}

\newcommand{\gaia}{{\it Gaia}}

\newcommand{\WASP}{{\it WASP-South}}
\newcommand{\wasp}{{\it WASP-South}}

\newcommand{\bagemass}{{\it BAGEMASS}}
\makeatletter
\newcommand\footnoteref[1]{\protected@xdef\@thefnmark{\ref{#1}}\@footnotemark}
\makeatother
\newcommand\B{\rule[-1.2ex]{0pt}{0pt}} 

\title[WASP-169, WASP-171, WASP-175 and WASP-182]{WASP-169, WASP-171, WASP-175 and WASP-182:   three hot Jupiters and one bloated sub-Saturn mass planet discovered by \wasp}

\author[L. D. Nielsen]{
L. D. Nielsen,$^{1}$\thanks{E-mail: Louise.Nielsen@unige.ch}
F. Bouchy,$^{1}$
O. D. Turner,$^{1}$
D.R. Anderson,$^{2}$
K. Barkaoui,$^{3,4}$
\newauthor
Z. Benkhaldoun$^{4}$
A. Burdanov,$^{3}$
A. Collier Cameron,$^{7}$
L. Delrez,$^{9,3}$
M. Gillon,$^{3}$
\newauthor
E. Ducrot,$^{3}$
C. Hellier,$^{2}$
E. Jehin,$^{3}$
M. Lendl,$^{1,8}$
P.F.L. Maxted,$^{2}$
F. Pepe,$^{1}$
D. Pollacco,$^{5,6}$
\newauthor
F.J. Pozuelos,$^{3}$
D. Queloz,$^{1,9}$
D. S\'egransan,$^{1}$
B. Smalley,$^{2}$
A.H.M.J. Triaud,$^{10}$
S. Udry,$^{1}$
\newauthor
and R.G. West$^{5,6}$ %
\\
$^{1}$Observatoire de Gen\`eve, Universit\'e de Gen\`eve, 51 Chemin des Maillettes, 1290 Sauverny, Switzerland\\
$^{2}$Astrophysics Group, Keele University, Staffordshire ST5 5BG, UK\\
$^{3}$Space sciences, Technologies and Astrophysics Research (STAR) Institute, Universit\'e de Li\`ege, Li\`ege 1, Belgium\\
$^{4}$Oukaimeden Observatory, High Energy Physics and Astrophysics Laboratory, Cadi Ayyad University, Marrakech, Morocco\\
$^{5}$Department of Physics, University of Warwick, Coventry CV4 7AL, UK\\
$^{6}$Centre for Exoplanets and Habitability, University of Warwick, Gibbet Hill Road, Coventry CV4 7AL, UK\\
$^{7}$SUPA, School of Physics and Astronomy, University of St. Andrews, North Haugh, Fife KY16 9SS, UK\\
$^{8}$Space Research Institute, Austrian Academy of Sciences, Schmiedlstr. 6, A-8042 Graz, Austria\\
$^{9}$Cavendish Laboratory, J J Thomson Avenue, Cambridge CB3 0HE, UK\\
$^{10}$School of Physics \& Astronomy, University of Birmingham, Edgbaston, Birmingham, B15 2TT, UK\\
}

\date{Accepted XXX. Received YYY; in original form ZZZ}

\pubyear{2019}

\begin{document}
\label{firstpage}
\pagerange{\pageref{firstpage}--\pageref{lastpage}}
\maketitle

\begin{abstract}
We present the discovery of four new giant planets from \WASP, three hot Jupiters and one bloated sub-Saturn mass planet; WASP-169b, WASP-171b, WASP-175b and WASP-182b. Besides the discovery photometry from \wasp\ we use radial velocity measurements from CORALIE and HARPS as well as follow-up photometry from EulerCam, TRAPPIST-North and -South and SPECULOOS.\\
WASP-169b is a low density Jupiter ($M=0.561 \pm 0.061~\mathrm{M_{Jup}}, R=1.304^{+0.150}_{-0.073} ~\mathrm{R_{Jup}}$) orbiting a V=12.17 F8 sub-giant in a 5.611~day orbit. \\
WASP-171b is a typical hot Jupiter ($M=1.084 \pm 0.094~\mathrm{M_{Jup}}, R=0.98^{+0.07}_{-0.04} ~\mathrm{R_{Jup}}$, $P=3.82~\mathrm{days}$) around a V=13.05 G0 star. We find a linear drift in the radial velocities of WASP-171 spanning 3.5 years, indicating the possibility of an additional outer planet or stellar companion.\\
WASP-175b is an inflated hot Jupiter ($M=0.99 \pm 0.13~\mathrm{M_{Jup}}, R=1.208 \pm 0.081 ~\mathrm{R_{Jup}}$, $P=3.07~\mathrm{days}$) around a V=12.04 F7 star, which possibly is part of a binary system with a star 7.9\arcsec\ away. \\
WASP-182b is a bloated sub-Saturn mass planet ($M=0.148 \pm 0.011~\mathrm{M_{Jup}}, R=0.850\pm 0.030 ~\mathrm{R_{Jup}}$) around a metal rich V=11.98 G5 star ([Fe/H]$=0.27 \pm 0.11$). With a orbital period of $P=3.377~\mathrm{days}$, it sits right in the apex of the sub-Jovian desert, bordering the upper- and lower edge of the desert in both the mass-period and radius-period plane.  \\
WASP-169b, WASP-175b and WASP-182b are promising targets for atmospheric characterisation through transmission spectroscopy, with expected transmission signals of 121, 150 and 264 ppm respectively.
\end{abstract}

\begin{keywords}
planets and satellites: detection -- planets and satellites: individual: WASP-169b -- planets and satellites: individual: WASP-171b -- planets and satellites: individual: WASP-175b -- planets and satellites: individual: WASP-182b 
\end{keywords}



\section{Introduction}
The Wide Angle Search for Planets (WASP; \citealt{2006PASP..118.1407P}) survey has since first light in 2006 discovered almost 200 transiting, close-in, giant exoplanets. These planets have provided great insight into exoplanetology as they enable studies of bulk properties, mass and radius from the transit photometry and radial velocity (RV) follow-up. Furthermore WASP have provided prime targets for in-depth characterisation of star-planet-interactions, exoplanet atmospheres \citep{2013MNRAS.436L..35B,2013A&A...554A..82D}, planetary winds \citep{2016ApJ...817..106B} and even the radial velocity shift of the planets themselves \citep{2010Natur.465.1049S}.


Furthermore, WASP and other wide field ground based surveys has been instrumental in discovering exoplanets bordering the sub-Jovian desert. The desert is constituted by a shortage of intermediate sized planets (1.0 - 0.1 $R_{\rm Jup}$) in close-in orbits (period < 5 days) \citep{Szabo2011,Mazeh2016,Fulton2018}. This phenomenon is evident when analysing the distribution of periods for exoplanets as a function of both planetary radius and mass, as illustrated in Fig. \ref{fig:MRP}. Ground based surveys have traditionally targeted planets sitting on the upper edge of the desert, due to detection limits. A notable exception to this rule is NGTS-4b \citep{2019MNRAS.486.5094W}; a sub-Neptune-sized planet in a 1.34 days orbit around a K-dwarf. NGTS-4b, is situated well within the sub-Jovian desert, challenging current theories of photo evaporation. 

Space based surveys, in particular \kepler\ \citep{Kepler}, have provided more targets constraining the lower edge, but mainly in the radius-period plane, as many of these targets are too faint for ground based follow up. The Transiting Exoplanet Survey Satellite, \citep[\tess,][]{Ricker2015}, is now changing the landscape of exoplanetology providing hundreds of transiting exoplanet candidates around bright stars, most of them appropriate for mass characterisation with RVs.

In this study we present four giant planets discovered with \wasp;
three hot Jupiters and one bloated sub-Saturn mass planet: WASP-169b, WASP-171b, WASP-175b and WASP-182b, all orbiting relatively bright G- and F-type stars. We perform a global MCMC analysis of the discovery data from \wasp, follow-up photometry from EulerCam, TRAPPIST-North, TRAPPIST-South and SPECULOOS and RVs from CORALIE and HARPS.
WASP-169b is a low density Jupiter ($M=0.561 \pm 0.061~\mathrm{M_{Jup}}, R=1.304^{+0.150}_{-0.073} ~\mathrm{R_{Jup}}, P=5.611~\mathrm{days}$) as well as WASP-175b ($M=0.99 \pm 0.13~\mathrm{M_{Jup}}, R=1.208 \pm 0.081 ~\mathrm{R_{Jup}}$, $P=3.07~\mathrm{days}$), making them interesting targets for atmopherics characterisation. WASP-171b is a typical hot Jupiter ($M=1.084 \pm 0.094~\mathrm{M_{Jup}}, R=0.98^{+0.07}_{-0.04} ~\mathrm{R_{Jup}}$, $P=3.82~\mathrm{days}$) with a possible additional companion indicated by a linear drift in the RVs. WASP-182b is a bloated sub-Saturn mass planet ($M=0.148 \pm 0.011~\mathrm{M_{Jup}}, R=0.850\pm 0.030 ~\mathrm{R_{Jup}}, P=3.377~\mathrm{days}$) sitting in the apex of the sub-Jovian desert, bordering the upper- and lower edge of the desert in both the radius-period and mass-period plane.

\section{Observations}
\subsection{Discovery photometry from WASP-south}
The host stars of the four planets presented in this paper have been surveyed by \wasp\ spanning several years, with WASP-182 being the target monitored for the longest time, dating back to 2006. WASP-South consisted, during the observations reported here, of eight 20 cm individual $f$/1.8 lenses mounted on the same fixture. Each lense was equipped with a 2k$x$2k CCD with a plate scale of 13.7\arcsec/pixel. The wide $7.8^{\circ} x 7.8^{\circ}$ field of view, allowed WASP-South to cover 1\% of the sky in each pointing, targeting stars with mV 9-13. The 20-cm lenses has since been replaced with 85 mm lenses, allowing the survey to target planets around brighter stars such WASP-189b \citep{Anderson2018}. 

Transit events are searched for in the discovery photometry using the box least square method as described in \cite{2006MNRAS.373..799C}. Targets with transits consistent with a planet-sized object are ranked according to \cite{2007MNRAS.380.1230C} and put forward for follow-up observations with a wide range of facilities. Both high resolution spectroscopy and photometry is used to confirm the planetary nature of the transiting object and ultimately measure both mass and radius precisely as described in the following sections. A summary of the observations used in this study can be found in Table \ref{tab:obs}.

\begin{table}
\caption{Summary of the discovery photometry, follow-up photometry and radial velocity observations of WASP-169, WASP-171, WASP-175 and WASP-182 from all facilities. Note time of meridian flip (MF) for WASP-169 on the TRAPPIST telescopes in BJD (\numprint{-2450000}).}
\label{tab:obs}
\begin{tabular}{llr}\hline
Date  & Source & N.Obs / Filter\\
\hline 
\multicolumn{1}{l}{{\bf WASP-169}} & & \\
2011 Jan--2012 Apr & WASP-South & \multicolumn{1}{r}{\numprint{24205}}  \\
2015 Mar--2017 May & CORALIE &  \multicolumn{1}{r}{25}  \\ 
2016 Feb 08 MF at 7427.6706 & TRAPPIST-South & I+z \\
2018 Jan 04 MF at 8123.5899 & TRAPPIST-North & I+z \\
2018 Dec 01 MF at 8454.6814 & TRAPPIST-North & z' \\
2019 Feb 23 MF at 8538.6286 & TRAPPIST-South & z' \\
\hline 
\multicolumn{1}{l}{{\bf WASP-171}} & \\
2011 Jan--2012 Jun & WASP-South & \multicolumn{1}{r}{\numprint{77507}}    \\ 
2015 Jun--2018 Dec & CORALIE &  \multicolumn{1}{r}{30}  \\ 
2018 May 15 & SPECULOOS-Io & I+z \\
\hline 
\multicolumn{1}{l}{{\bf WASP-175}} & \\
2013 Jan--2014 Jun & WASP-South & \multicolumn{1}{r}{\numprint{86025}}   \\ 
2015 Jun--2018 Jul & CORALIE &  \multicolumn{1}{r}{20}   \\ 
2014 Apr 15 & TRAPPIST-South & Blue Blocking \\ 
2015 Dec 19 & TRAPPIST-South & Blue Blocking \\ 
2016 Dec 30 & EulerCam & BG \\
2017 Feb 11 & TRAPPIST-South & z' \\ 
\hline 
\multicolumn{1}{l}{{\bf WASP-182}} & \\
2006 May--2014 Nov & WASP-South & \multicolumn{1}{r}{\numprint{127127}}   \\
2016 June--2018 Jul & CORALIE &  \multicolumn{1}{r}{21}   \\
2018 Mar--2018 Nov & HARPS &  \multicolumn{1}{r}{14}    \\
2015 Oct 23 & TRAPPIST-South & I+z \\
2018 Jun 28 & Euler Cam & RG \\
2018 Aug 01 & Euler Cam & RG \\
2018 Aug 11 & TRAPPIST-South & I+z \\
2018 Aug 28 & TRAPPIST-South & I+z \\
\hline 
\end{tabular}
\end{table}

\subsection{CORALIE spectroscopy}
{Several spectra at different epochs} were obtained for all four targets using the high resolution spectrograph CORALIE on the Swiss 1.2-m Euler telescope at \LSO, Chile \citep{CORALIE}. CORALIE has a resolving power of $R\sim\numprint{60000}$ and is fed by two fibres; one 2\arcsec\ on-sky science fibre encompassing the star and another which can either be connected to a Fabry-P\'{e}rot etalon for simultaneous wavelength calibration or on-sky for background subtraction of the sky-flux. For WASP-169, WASP-171 and WASP-175 the CORALIE spectra were used to derive stellar parameters, see Sec. \ref{sec:spectra} for a detailed description of the analysis. 

\begin{figure}
	\includegraphics[width=0.99\columnwidth,trim={0cm 0cm 0.0cm 0cm},clip]{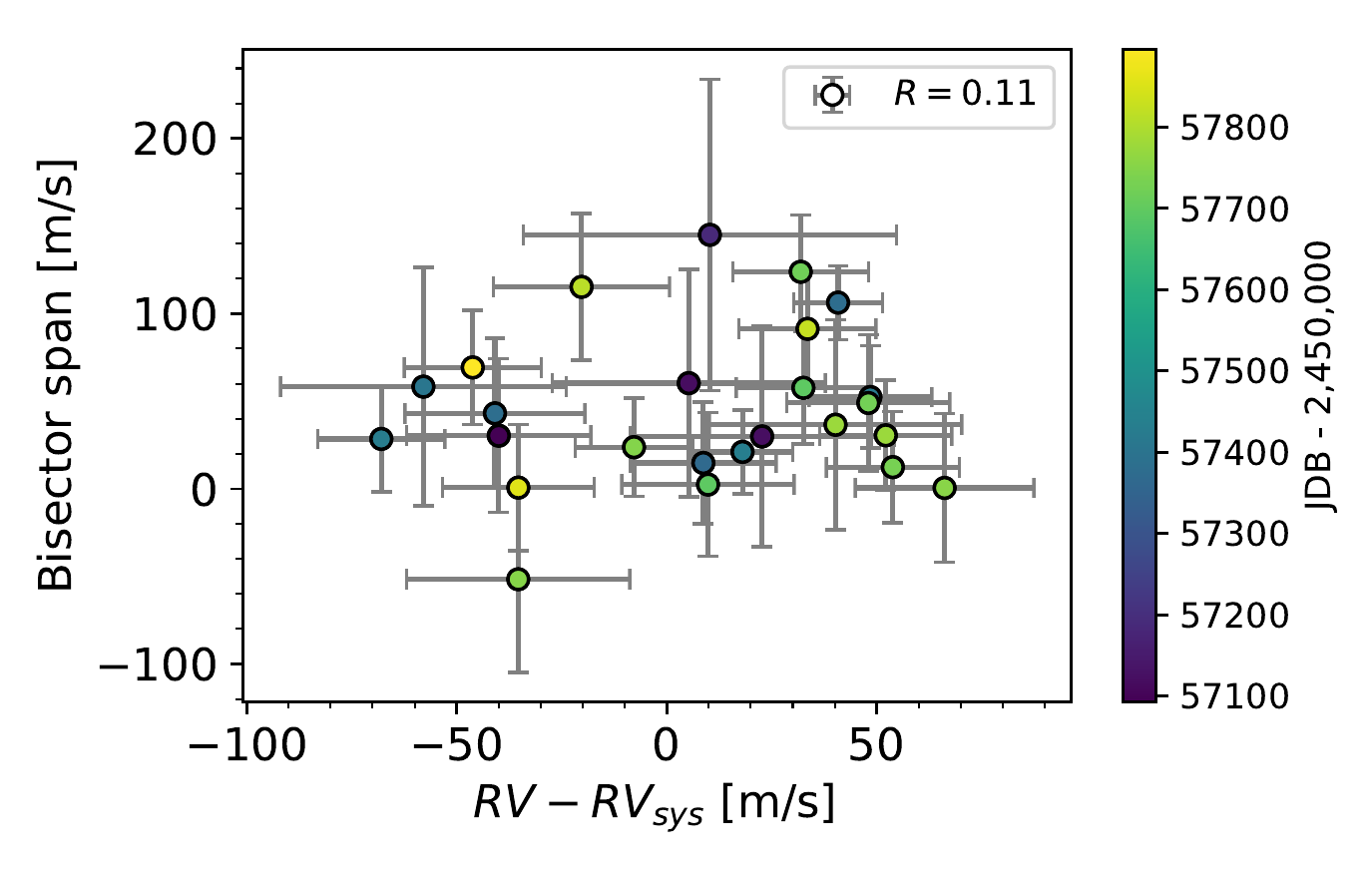}
	\includegraphics[width=0.99\columnwidth,trim={0.0cm 0cm 0.0cm 0cm},clip]{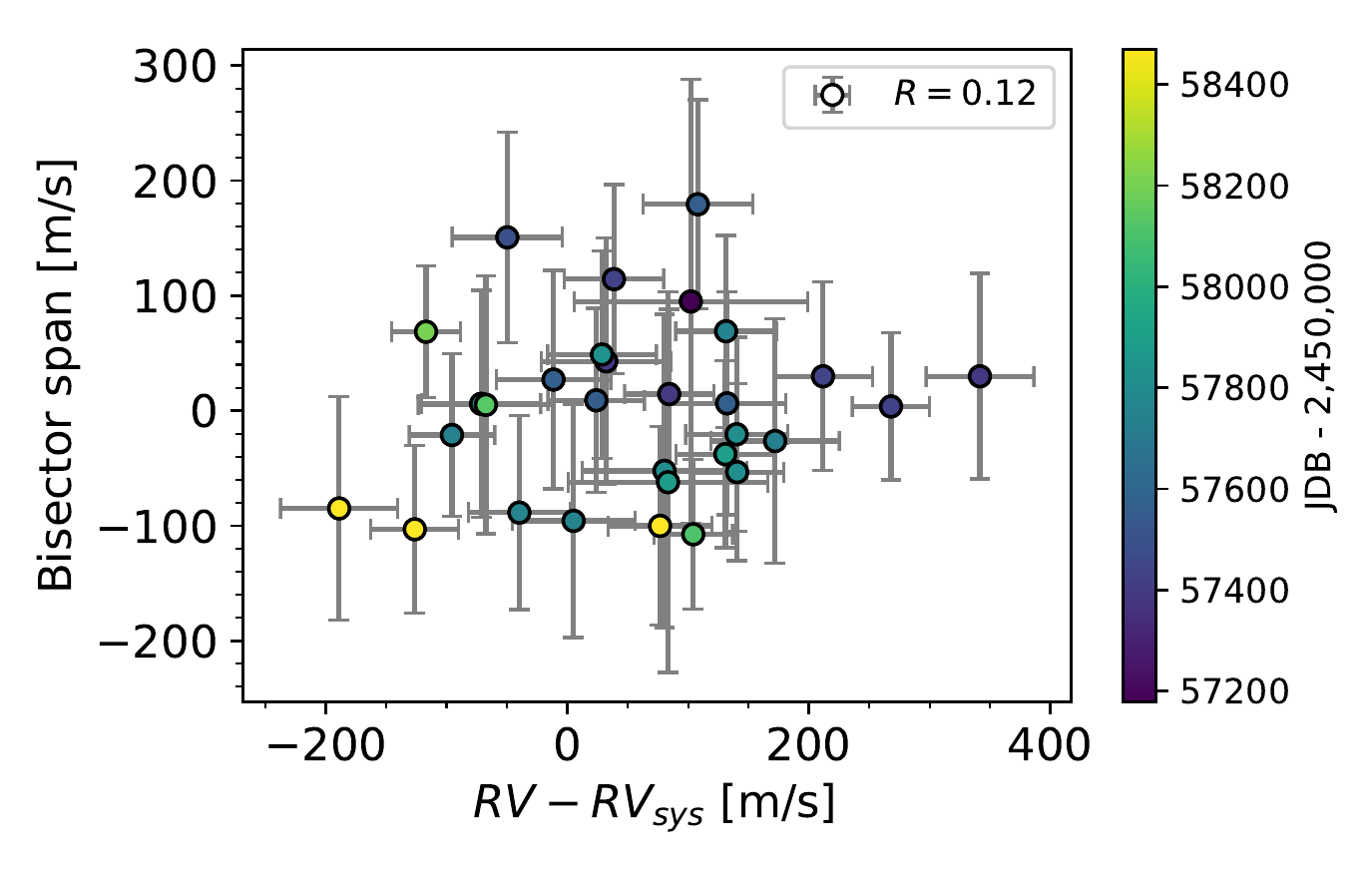}
	\includegraphics[width=0.99\columnwidth,trim={0cm 0cm 0.0cm 0cm},clip]{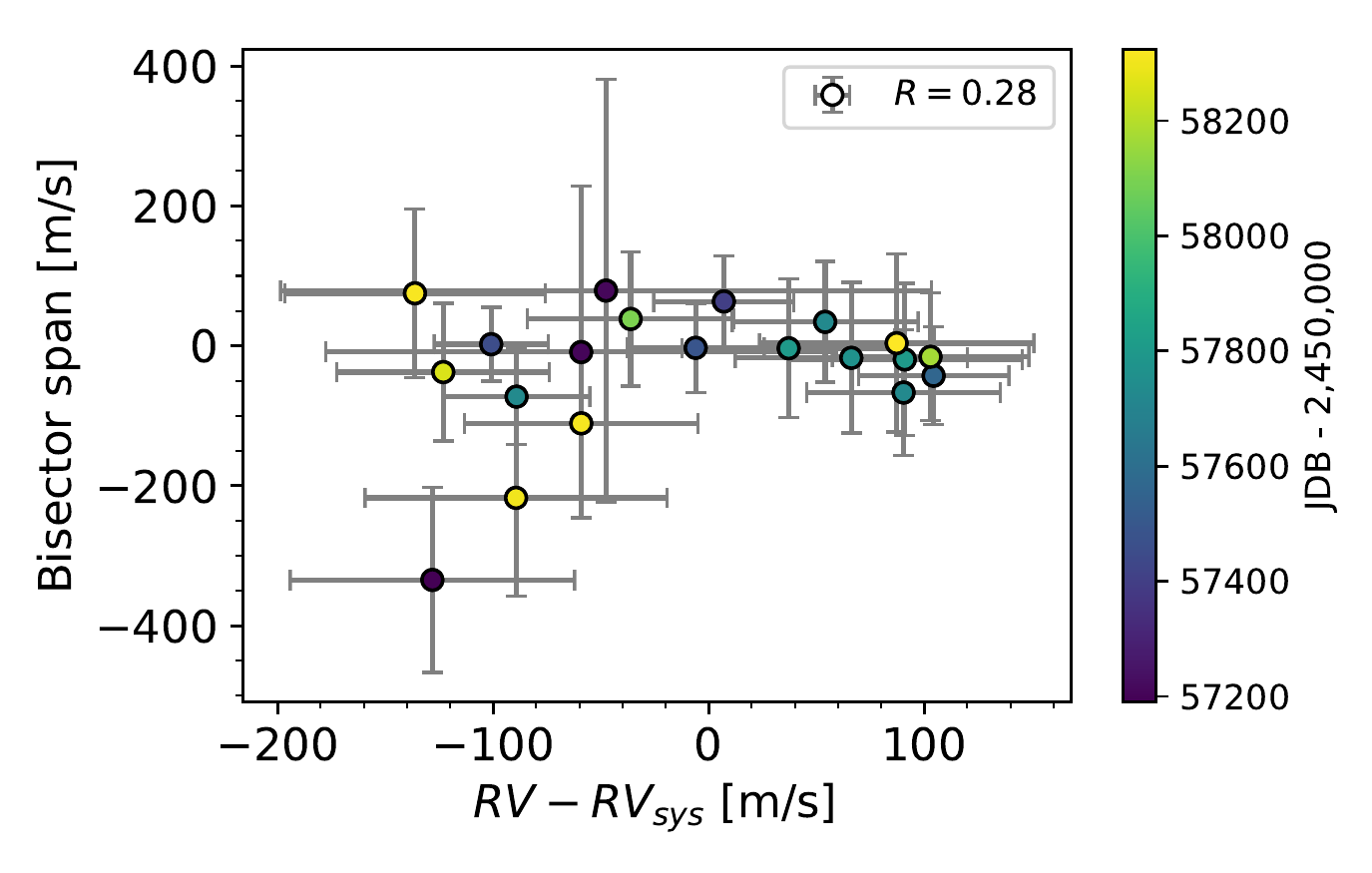}
	\includegraphics[width=0.99\columnwidth,trim={0.0cm 0cm 0.0cm 0cm},clip]{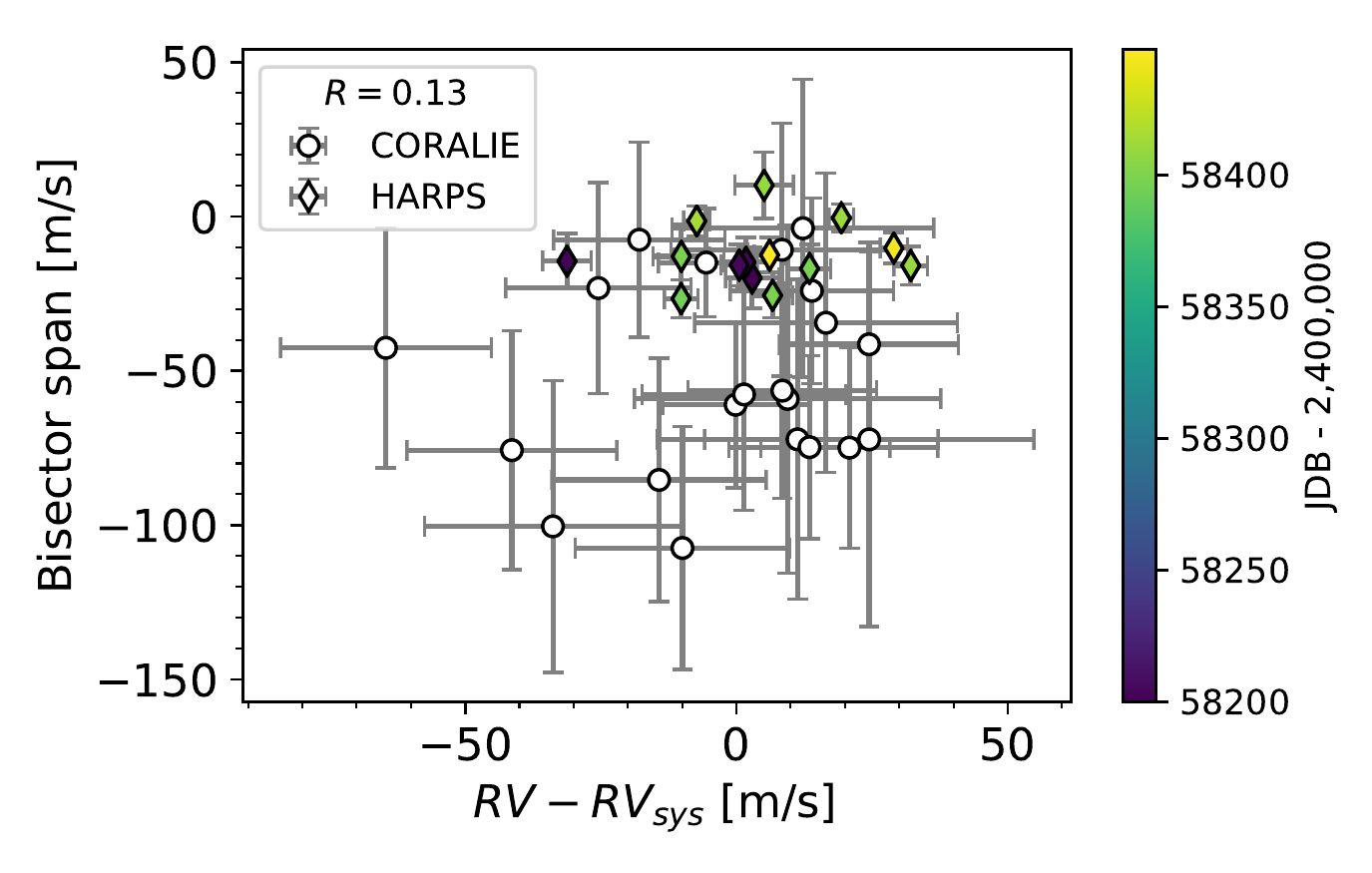}
    \caption{Bisector-span and RVs for WASP-169, WASP-171, WASP-175 and WASP-182 from top to bottom panel. The Pearson coefficient $R$ shows there are no correlation between the bisector-span and RVs.  For WASP-182 no offset in bisector span from CORALIE to HARPS were corrected for, and the Pearson coefficient is for the HARPS RVs only.}
    \label{fig:bis}
\end{figure}

We obtained RVs for each epoch by cross-correlating with a binary G2 mask \citep{Pepe2002}. Bisector-span, FWHM and other line-profile diagnostics were computed as well. Figure \ref{fig:bis} shows RVs and bisector span for the four stars, with Pearson-coefficients. No correlation was found between the RVs and the bisector-span. We also computed RVs using other binary masks ranging from A0 to M4, to check for any mask-dependent signal indicating a blend. As such, the CORALIE RVs confirm the planetary nature of the transit signals and we found them all to be in phase with the transit events detected by \wasp.

Table \ref{tab:rvs} show the 5 first RVs of WASP-169 from CORALIE, along with RV uncertainty, FWHM of the CCF and bisector span. Full ascii tabels with all the RV data presented in this study are available online.

\subsection{HARPS spectroscopy}
To enable precise mass measurement of WASP-182b we also obtained HARPS RVs under programmes Anderson: 0100.C-0847 and Nielsen: 0102.C-0414 in 2018. HARPS is hosted by the ESO 3.6-m telescope at \LSO, Chile \citep{2003Msngr.114...20M} and has a resolving power of $R\sim \numprint{100000}$. The RVs were computed using the standard data-reduction pipeline with a binary G2 mask, and confirmed the RV-amplitude found with CORALIE, though with greater precision. The HARPS spectra were also used to derive spectral parameters for WASP-182, as detailed in Sec. \ref{sec:spectra}.

\begin{table}
\caption{The first five radial velocity measurements for WASP-169 from CORALIE, along with RV uncertainties, $\sigma_{RV}$, FWHM of the CCF and bisector-spans. BJD is barycentric Jullian dates. Full machine-readable tables for all four stars are available with the online journal.}
\label{tab:rvs}
\begin{tabular}{ccccc}\hline
Time  & RV & $\sigma_{RV}$ & FWHM  & Bisector\\
(BJD - \numprint{2400000}) & $(\textrm{km s}^{-1})$ &$(\textrm{km s}^{-1})$ & $(\textrm{km s}^{-1})$  & $(\textrm{km s}^{-1})$\\
\hline 
57092.695347 & 67.61532 & 0.02192 & 10.45130	& 0.03031 \\
57119.639402 & 67.66054 & 0.03251 & 10.57522	& 0.06030 \\
57121.569521 & 67.67801 & 0.03145 & 10.42107	& 0.02989 \\
57185.451344 & 67.66557 & 0.04442 & 10.53662	& 0.14495 \\
57365.775616 & 67.66402 & 0.01734 & 10.41273	& 0.01473 \\
...&...&...&...&...\\
\hline 
\end{tabular}
\end{table}

\subsection{EulerCam}
Additional photometry was acquired for WASP-175 and WASP-182 using EulerCam \citep{2012A&A...544A..72L}, also on the 1.2-m Swiss at \LSO. The observations used B and R filters, respectively. The data were bias and flat field corrected and photometry extracted for a number of comparison stars and aperture radii. The comparison star ensemble and aperture radii chosen such that the scatter in a simple linear fit to the out of transit portion was minimised. The aim of this process was to produce a final light curve optimised to reduce the overall scatter.

\subsection{TRAPPIST-North and -South}
Both of the two 0.6-m TRAPPIST telescopes \citep{2011A&A...533A..88G,2011Msngr.145....2J}, based at La Silla and Oukaimeden Observatory in Morocco \citep{Gillon2017Natur,Barkaoui2019} were used to perform follow-up photometry on WASP-169, WASP-175 and WASP-182. All light curves of WASP-169 contain a meridian flip (MF), as detailed in Table \ref{tab:obs}. In the joint analysis of the RVs and photometry, described in Section \ref{sec:mcmc}, the data were partitioned at the time of MF and modelled as two independent data sets.


Data reduction consisted of standard calibration steps (bias, dark and flat-field corrections) and subsequent aperture photometry using IRAF/DAOPHOT \citep{Tody1986}. Extraction of fluxes of selected stars using aperture photometry was performed with IRAF/DAOPHOT, as described in \cite{Gillon2013}.

\subsection{SPECULOOS-South}
The robotic 1-m SSO-Io telescope is one of four telescopes at the SPECULOOS-South facility located at \PAR, Chile \citep{2018Msngr.174....2J, Delrez2018, Gillon2018, Burdanov2017}. It started its science operations in 2017 and observed one full transit of WASP-171 in May 2018 using a I+z filter, toward the near-infrared end of the visible spectrum. The SPECULOOS telescopes are equipped with 2K$x$2K CCD cameras, with increased sensitivities up to 1 \micron, in the very-near-infrared. The calibration and photometric reduction of the data were performed as described in \cite{Gillon2013}.  

\section{Stellar parameters} \label{sec:star}

\subsection{Spectral characterisation} \label{sec:spectra}
Following the methods described in \cite{Doyle2012} we used the CORALIE and HARPS spectra to derive stellar parameters. Effective temperature, $T_{\rm eff}$, is computed from the $\mathrm{H \alpha}$-line. Surface gravity, $\log g$, is based on Na I D and Mg I b lines. The metallicity, [Fe/H], is determined from the equivalent-width of a selection of unblended Fe-lines. Lithium abundances which can be used to gauge stellar age and has been proposed to be a tracer of planet formation \citep{King1997, Figueira2014}, are derived as well. The uncertainty on $T_{\rm eff}$ and $\log g$ is propagated through to the abundances. 

The projected rotational velocity, $V \sin i$, is found by convoluting the width of stellar absorption lines with the instrumental resolution ($\mathrm{R \sim \numprint{60000}}$ for CORALIE and $\mathrm{R \sim \numprint{100000}}$ for HARPS) and modelling macro turbulence by the method proposed in \cite{2014MNRAS.444.3592D}. Micro turbulence was estimated using the calibration from \cite{2012MNRAS.423..122B}.

\subsection{Stellar masses and ages with \bagemass} \label{sec:bagemass}
\begin{figure}
	\includegraphics[width=\columnwidth,trim={0.5cm 0.0cm 1cm 1cm},clip]{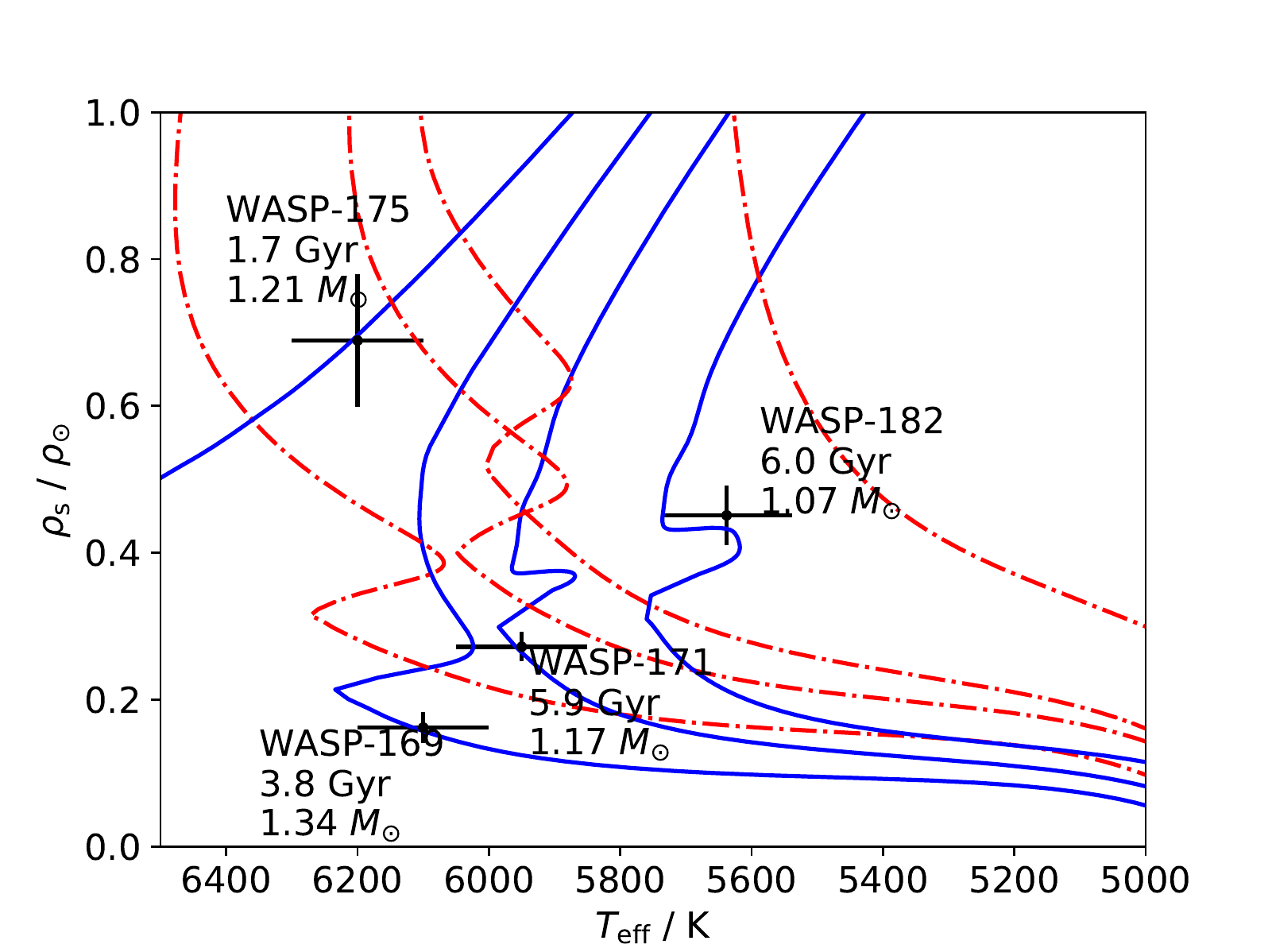}
    \caption{Isochrones (solid/blue) and evolution tracks (dot-dashed/red) output by \bagemass\ for each of the four host stars (labelled).}
    \label{fig:tracks}
\end{figure}

We used the Bayesian stellar evolution code \bagemass\ \citep{bagemass} to model stellar masses and ages based on spectral $T_{\rm eff}$ and [Fe/H] as well as the stellar density derived from the transit-light curves. \bagemass\ samples a dense grid of stellar models to compute stellar masses and ages. The stellar masses obtained were used as Gaussian inputs in the final joint model. Figure \ref{fig:tracks} shows the stellar evolutionary tracks and isochrones for all four planet host stars. All adopted stellar parameters from spectral characterisation, \bagemass, and the final joint model are listed in Table \ref{tab:params}.

\subsection{Rotational modulation} \label{sec:rot}
We searched for rotational modulation caused by stellar spots in the \WASP\ light curves for the four host stars using the method described in \cite{2011PASP..123..547M}. Star spots have limited lifetimes and will have variable distribution on the stellar surface over time. Therefore the modulation is not expected to be coherent, and so we  searched each season of \wasp\ data individually. WASP-169 and WASP-171 showed no significant modulation, with an upper limit on the amplitude of 1.5~mmag. For WASP-175 we can set an upper limit of 2~mmag. 

\begin{figure}
    \includegraphics[width=\columnwidth,trim={0.5cm 0.0cm 0.2cm 0.0cm},clip]{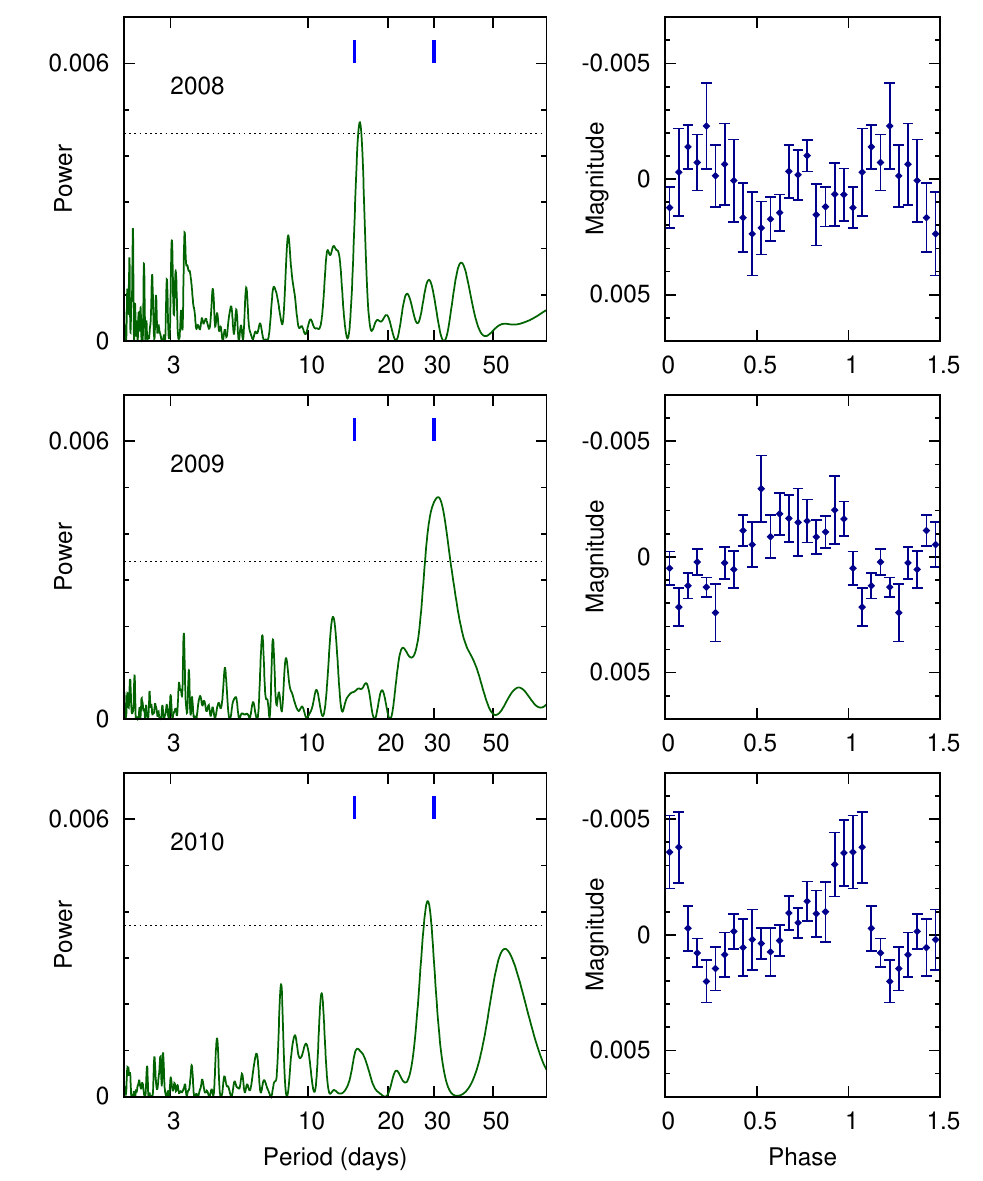}
    \caption{\label{fig:rotW182}Periodograms of the \wasp\ data for WASP-182 from three different years (left) along with folds of the data at the possible rotational periods (right). The folded data has been binned to 20 bins, each corresponding to 1.5 nights.  The blue marks are at 30 and 15 days and the dotted horizontal line at 1\% FAP.}
\end{figure}

For WASP-182 we find a possible modulation in the data from both 2009 and 2010, with a false-alarm probability of 1\%\ in each case. The modulation has a period of 30 +/- 2 days and an amplitude of 1 to 2 mmag, which is near the detection limit in \wasp\ data. In 2008 we saw a peak near (but not exactly at) half the period seen in 2009 and 2010, which could thus be the first harmonic of the rotational modulation (see Fig.~\ref{fig:rotW182}). The exact position of the strongest peak in the data from 2009 and 2010 differ slightly. This could be a result of us tracking star spots at different latitudes on the stellar disk between the two seasons, which in the presence of differential rotation will cause a phase shift in modulation. Another possibility is star-spot groups coming and going during the season, which also will induce a period shift in the periodogram.
In data both before (2006 and 2007) and after (2011 and 2012) these years we see no significant modulation, though in each case the data are less extensive than in 2009 and 2010. 
The 30-day rotation period corresponds to a rotational velocity of about 2 km/s and is consistent with the $ V \sin i$ computed from HARPS spectra ($1.4 \pm 1.0$~km/s).

\section{System parameters} \label{sec:mcmc}
The full set of system parameters were modelled jointly using the discovery photometry, follow-up light curves and RV data with the Markov-Chain Monte Carlo (MCMC) code described in detail in \cite{2007MNRAS.380.1230C} and \cite{2015A&A...575A..61A}. The analytic eclipse-expressions derived by \cite{2002ApJ...580L.171M} are used with a 4 parameters, non-linear limb darkening law of \cite{2000A&A...363.1081C,2004A&A...428.1001C}. We have interpolated coefficients for stellar temperature and metallicity of each star, and in each photometric filter. The values used were perturbed during the MCMC via $T_{\rm L-D}$, the `limb-darkening temperature', which has a mean and standard deviation corresponding to the spectroscopic $T_{\rm eff}$ and its uncertainty. 

We ran the MCMC both with the eccentricity as a free parameter and fixed to zero, to check if the results are compatible with a circular orbit. We expect most giant planets in short period orbits to have been circularised by tidal forces, and want to avoid over-estimating the eccentricity in orbits that have none. 
Each circular model has 6 fitted parameters; orbital period , $P$, epoch, $T_{\rm C}$, transit depth in the absences of dark limb effects, $(R_{p}/R_{s})^{2}$, transit duration, $T_{14}$, impact parameters, $b$ and stellar radial velocity semi-amplitude, $K_1$. The RV systemic velocity $\gamma$ was fitted too, and in the case of WASP-171 along with a linear RV-drift, $\dot \gamma$. For WASP-182, where we have data from two different spectrographs, an offset between CORALIE and HARPS was modelled as well. 

For each target we ran 5000 MCMC steps as a 'burn in phase' to initialise the main phase which was set to have \numprint{50000} iterations. At each step the free parameters are perturbed and the models are re-fit. If the $\chi^2$ of the fit is better than the previous step the current parameters are accepted, if the fit is worse the parameters are accepted with a probability proportional to $\exp(-\Delta\chi^2)$. We used Gelman-Rubin statistics \citep{Gelman2003, ford2006} to check how well the chains converge. In our case the Gelman-Rubin statistics indicated that all fitted parameters were well mixed. 

Continuing our practice from recent discovery papers \citep[e.g.][]{Hellier2019} we treat the stellar parameters through a two-step process; we first estimate the stellar density, $\rho_S$, from the transit duration alone, independently of stellar models. Secondly we obtain stellar masses by using $\rho_S$, $T_{\rm eff}$ and [Fe/H] in the stellar evolution model \bagemass, as explained in Section \ref{sec:bagemass}. The resulting stellar mass estimate and its uncertainty is finally used as input in the MCMC to derive stellar radii. The stellar density for WASP-182 is poorly constrained by the transit data alone, so we used an additional prior on the radius from \gaia\ DR2, as described in \cite{Turner2019}.

\section{Results}
For each system we list the final stellar and planetary parameters in Table \ref{tab:params} with 1-$\sigma$ errors. Figures \ref{fig:W169} through \ref{fig:W182} show the final joint model fitted to the discovery and follow-up data.


\begin{table*}
\begin{minipage}{\textwidth}
\caption{System parameters for WASP-169, WASP-171, WASP-175 and WASP-182, based on the analysis presented in Section \ref{sec:star} \& \ref{sec:mcmc}.  Adopted non SI-units are $\mathrm{M_{\odot}} = 1.9891\cdot 10^{30} ~\mathrm{kg}$, $\mathrm{R_{\odot}} = 6.95508 \cdot 10^{8}~ \mathrm{m}$, $\mathrm{R_{Jup}} = 7.149253763\cdot 10^{7}~ \mathrm{m}$ and $\mathrm{M_{Jup}} =\mathrm{M_{\odot}} / 1047.52 $.}
\label{tab:params}
\begin{adjustbox}{width=\textwidth}
\begin{tabular}{lccccc}\hline
Parameter & Symbol (Unit) & WASP-169  & WASP-171 & WASP-175 & WASP-182\\ 
\hline \hline
\textbf{Stellar parameters} &&&&&\\
~~~\WASP\ ID & 1SWASPJ... & 082932.97-125640.9 & 112722.86-440519.3 & 110516.60-340720.3 &204641.58-414915.2 \\
~~~{\it 2MASS} ID\footnote{\cite{2MASS}} &   & J08293295-1256411    & J11272283-4405193  & J11051653-3407219 & J20464156-4149151\\
~~~Right ascension      & RA (hh:mm:ss) & 08:29:32.97  & 11:27:22.86 & 11:05:16.60 &  20:46:41.58\\
~~~Declination    & DEC (dd:mm:ss) & -12:56:40.9  & -44:05:19.3 & -34:07:20.3 &  -41:49:15.2\\
~~~Visual magnitude\footnote{From NOMAD \citep{NOMAD}} & mV (mag) &  12.17 & 13.05 & 12.04  &  11.98 \\
~~~Stellar Mass & $M_{s}$ $\mathrm{(M_{\odot})}$ & $1.337 \pm 0.083$ & $1.171 \pm 0.058$ & $1.212 \pm 0.045$ & $1.076 \pm 0.064$ \\ 
~~~Stellar Radius & $R_{s}$ $\mathrm{(R_{\odot})}$ &  $2.011^{+0.188}_{-0.089}$ &  $1.637^{+0.091}_{-0.046}$ & $1.204 \pm 0.064$ & $1.34 \pm 0.03$ \\ 
~~~Effective temp.\footnote{\label{spec}From spectral analysis of CORALIE spectra (HARPS for WASP-182 (Sec. \ref{sec:spectra})}
& $T_{\textrm{eff}}$ (K) & $6110 \pm 101$ & $5965 \pm 100$ & $6229 \pm 100$ & $5638 \pm 100 $ \\ 
~~~Stellar metallicity \footnoteref{spec}
& ${[}\textrm{Fe/H}{]}$ & $0.06 \pm 0.07$ & $0.04 \pm 0.07$ & $0.150 \pm 0.069$ & $0.27 \pm 0.11$ \\ 
~~~Lithium abundance\footnoteref{spec} & $\log {\rm A(Li)} $ & None found &$\sim 1.1 \pm 0.2$ & $2.16 \pm 0.08 $& $2.0 \pm 0.09$\\
~~~Macro-turbulent vel.\footnote{\label{turb} Derived via the method by \cite{2014MNRAS.444.3592D} on CORALIE and HARPS spectra}
& $ V_{\rm mac} (\textrm{km s}^{-1})$& 5.0 & 4.4 & $\le 4.8$ & $3.4 \pm 0.7$ \\
~~~Projected rot. vel.\footnote{Derived from CORALIE and HARPS spectra, assuming Macro-turbulent velocity\footnoteref{turb}} & $V\sin i$  $(\textrm{km s}^{-1})$ & $4.3 \pm 0.9$ & $6.3 \pm 0.9$ & $\le 4.0$ & $1.4 \pm 1.0$ \\ 
~~~Age\footnote{From \bagemass\ analysis (Sec. \ref{sec:bagemass})} & (Gyr) & $3.802  \pm 0.779 $ & $5.908 \pm 1.051 $ & $1.745 \pm  0.995$ & $5.952 \pm 2.684$ \\
~~~Distance\footnote{From \gaia\ DR2 parallax \citep{GAIA2016,GAIADR2}} & $d$~(pc) & $638 \pm 14$ & $774 \pm 20$ & $584 \pm 13$ & $331 \pm 4.6$ \\
~~~Stellar density & $\rho_{s}$ ($\rho_{\odot}$)  &  $0.166^{+0.019}_{-0.039}$ &  $0.270^{+0.017}_{-0.042}$ &  $0.693^{+0.125}_{-0.094}$ & $0.451 \pm 0.041$ \\ 
~~~Surface gravity & $\log(g_{s})$ (cgs) &  $3.958^{+0.033}_{-0.076}$ &  $4.080^{+0.020}_{-0.049}$ & $4.359 \pm 0.045$ & $4.218 \pm 0.033$ \B \\ 
\hline  
\textbf{Planet parameters} &&&&& \\
~~~Planet mass & $M_{p}$ $(\mathrm{M_{Jup}})$ & $0.561 \pm 0.061$ & $1.084 \pm 0.094$ & $0.99 \pm 0.13$ & $0.148 \pm 0.011$ \\ 
~~~Planet radius & $R_{p}$ $(\mathrm{R_{Jup}})$ &  $1.304^{+0.150}_{-0.073}$ &  $0.98^{+0.07}_{-0.04}$ & $1.208 \pm 0.081$ & $0.850 \pm 0.030$ \\ 
~~~Period & $P$ (d) & $5.6114118 \pm 0.0000092$ & $3.8186244 \pm 0.0000038$ &  $3.0652907^{+0.0000011}_{-0.0000016}$ & $3.3769848 \pm 0.0000024$ \\ 
~~~Transit epoch & $T_{\mathrm{C}} - \numprint{2400000} $  & $57697.0176 \pm 0.0014$ & $58059.8295 \pm 0.0011$ & $57143.78886 \pm 0.00034$ & $58018.66018 \pm 0.00067$ \\ 
~~~Transit duration & $T_{14}$ (d) & $0.2522^{+0.0042}_{-0.0034}$ & $0.1908 \pm 0.0024$ & $0.1115 \pm 0.0017$ & $0.1082 \pm 0.0015$ \\ 
~~~Transit depth & $(R_{p}/R_{s})^{2}$ & $0.00446 \pm 0.00028$ & $0.00382 \pm 0.00018$ & $0.01064 \pm 0.00036$ & $0.00426 \pm 0.00017$ \\ 
~~~Scaled semi-major axis & $a/R_{s}$ &  $7.30^{+0.68}_{-0.26}$ &  $6.64^{+0.38}_{-0.13}$ & $7.86 \pm 0.41$ &  $8.3^{+1.9}_{-1.4}$ \\ 
~~~Semi-major axis & $a$ (au) &  $0.0681 \pm 0.0014$ & $0.05040 \pm 0.00083$ & $0.04403 \pm 0.00055$ & $0.0451 \pm 0.0009$ \\ 
~~~Impact parameter & $b$ & $0.27 \pm 0.19$ &  $0.19^{+0.19}_{-0.13}$ &  $0.640^{+0.043}_{-0.061}$ & $0.775 \pm 0.019$ \\ 
~~~Orbital eccentricity & $e$ & 0 (adopted, $2\sigma< 0.17$) & 0 (adopted, $2\sigma< 0.16$)& 0 (adopted, $2\sigma< 0.28$) & 0 (adopted, $2\sigma< 0.25$) \\ 
~~~Orbital inclination & $i$ ($\degr$) &  $87.9^{+1.4}_{-2.0}$ &  $88.3^{+1.1}_{-1.9}$ & $85.33 \pm 0.62$ & $83.88 \pm 0.33$ \\ 
~~~RV Semi-amplitude & $K_1~(\rm \textrm{m s}^{-1})$ & $52.9 \pm 5.4$ & $126 \pm 10$ & $124 \pm 17$ & $19.0 \pm 1.2$ \\ 
~~~Systemic RV & $\gamma$~$(\textrm{km s}^{-1})$ & $67.6553 \pm 0.0035$ & $10.4838 \pm 0.0078$ & $5.38 \pm 0.01$ & $-34.1325 \pm 0.0038$ \footnote{RV offset of $31.9 \pm 0.2\ \textrm{m s}^{-1}$ from CORALIE to HARPS is found.} \\ 
~~~RV drift & $ \dot \gamma$ $(\textrm{km s}^{-1}\textrm{yr}^{-1})$ & -  & $-0.07659 \pm  0.0085$ & - & - \\ 
~~~RV residuals & $ \sigma (res_{\textrm{RV}})$ $(\textrm{m s}^{-1})$ & 18  & 34 & 6.5\footnote{RV residual for HARPS and CORALIE combined.} & 34 \\ 
~~~Reduced RV fit $\chi^2$ & $ {\chi}^2_{r}$ & 1.2 & 1.1 & 1.9 & 0.43 \\ 
~~~Planet density & $\rho_{p}$ ($\rho_{\textrm{Jup}})$ &  $0.249^{+0.056}_{-0.071}$ &  $1.13^{+0.17}_{-0.22}$ &  $0.56^{+0.15}_{-0.11}$ & $0.240 \pm 0.044$ \\
~~~Surface gravity & $\log(g_{p})$ (cgs) &  $2.87^{+0.07}_{-0.10}$ &  $3.405^{+0.049}_{-0.069}$ & $3.194 \pm 0.077$ & $2.669 \pm 0.049$ \\ 
~~~Equilibrium temperature & $T_{\mathrm{eq}}$ (K) &  $1604^{+74}_{-42}$ &  $1642^{+51}_{-35}$ & $1571 \pm 49$ & $1479 \pm 34$ \B \\ \hline
\end{tabular}
\end{adjustbox}
\end{minipage}
\end{table*}

\subsection{WASP-169b}
\begin{figure}
	\includegraphics[width=\columnwidth]{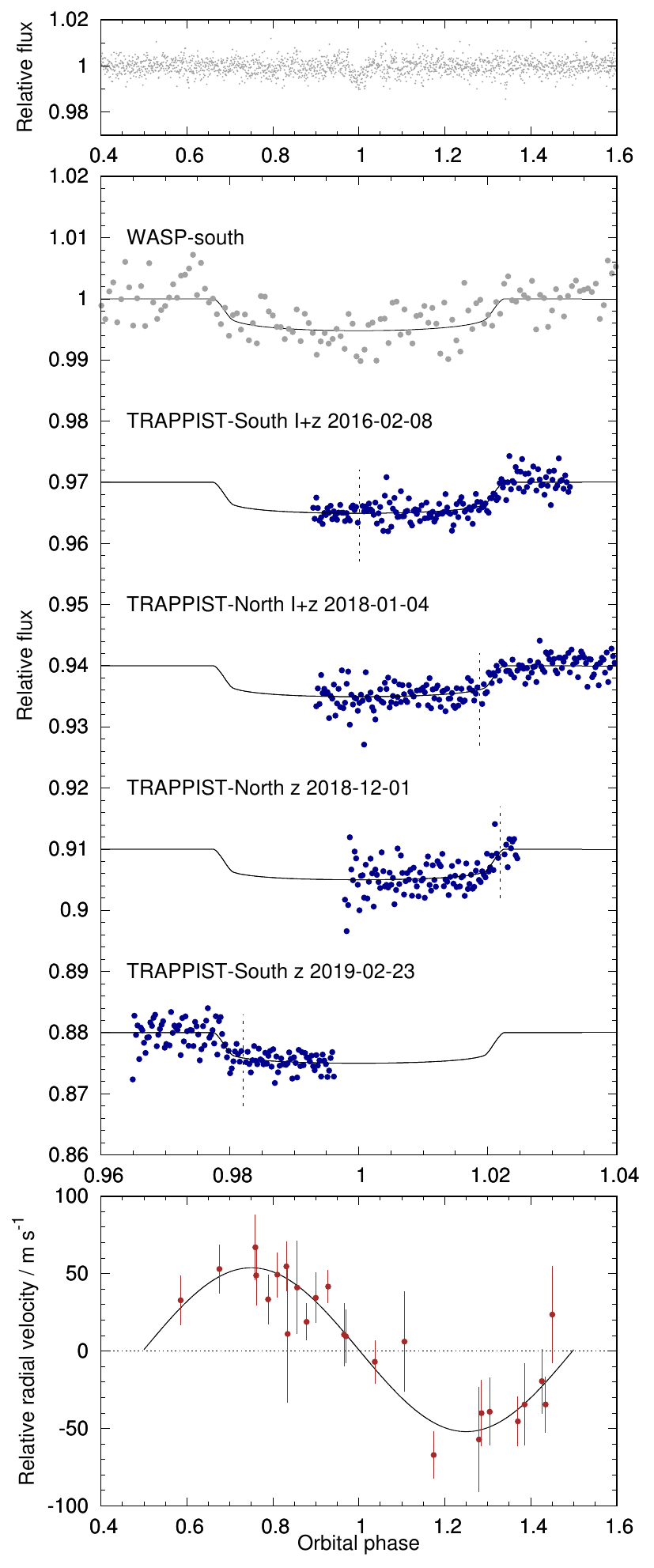}
    \caption{Data for the WASP-169 system. {\it Top}: WASP discovery light curve phase-folded on period found by joint analysis and binned to 2 minutes. {\it Middle}: Light curves used in joint analysis. The WASP light curve has been binned to 5 minutes and is shown as grey points with the transit model overplotted. The follow-up light curves have been binned to 2 minutes and are here all from TRAPPIST-North and South shown in blue. Times of meridian flip are indicated as vertical dashed lines. {\it Bottom}: CORALIE radial velocities used in the joint analysis over plotted with resulting model.}
    \label{fig:W169}
\end{figure}

WASP-169b is a low density Jupiter with mass $0.561 \pm 0.061~\mathrm{M_{Jup}}$ and radius $1.304^{+0.150}_{-0.073} ~\mathrm{R_{Jup}}$ in a 5.611 day orbit around a V=12.17 F8 sub-giant. Figure \ref{fig:W169} shows the \wasp\ discovery light curve with follow-up observations from TRAPPIST-North, -South and CORALIE. The planetary and stellar parameters are well constrained. The transit $\log(g_{s})=3.958^{+0.033}_{-0.076}$ (cgs) is consistent with the spectroscopic value of $4.0 \pm 0.2$. The resulting stellar radius ($2.011^{+0.188}_{-0.089} \mathrm{R_{\odot}}$) is in agreement with \gaia\ DR2 ($2.28^{+0.10}_{-0.25} \mathrm{R_{\odot}}$). WASP-169 has a faint star 7\arcsec\ away with $\Delta$g=5.4 \citep{GAIADR2}. It has a similar parallax ($1.26\pm 0.09$ mas vs $1.566 \pm 0.04$ mas), but does not appear to be co-moving.

The low density of WASP-169b ($0.249 ^{+0.056}_{-0.071} \mathrm{\rho_{jup}}$) should make it a good candidate for atmospheric characterisation. It has an estimated scale height of 1300 km, corresponding to a transmission signal of 121 ppm. The JWST instrument NIRSpec will uniquely be able to cover the near-infrared spectral range from 0.6 to 5.3 \micron\ in one low resolution spectrum in 'PRISM mode' \citep{Birkmann2016}. With a J-band magnitude of 10.8, WASP-169 is a perfect target for NIRSpec, expecting to achieving SNR \numprint{10000} - \numprint{25000} per resolution element across the spectrum with one transit \citep{Nielsen2016, pandexo}.

\subsection{WASP-171b}
\begin{figure}
	\includegraphics[width=\columnwidth]{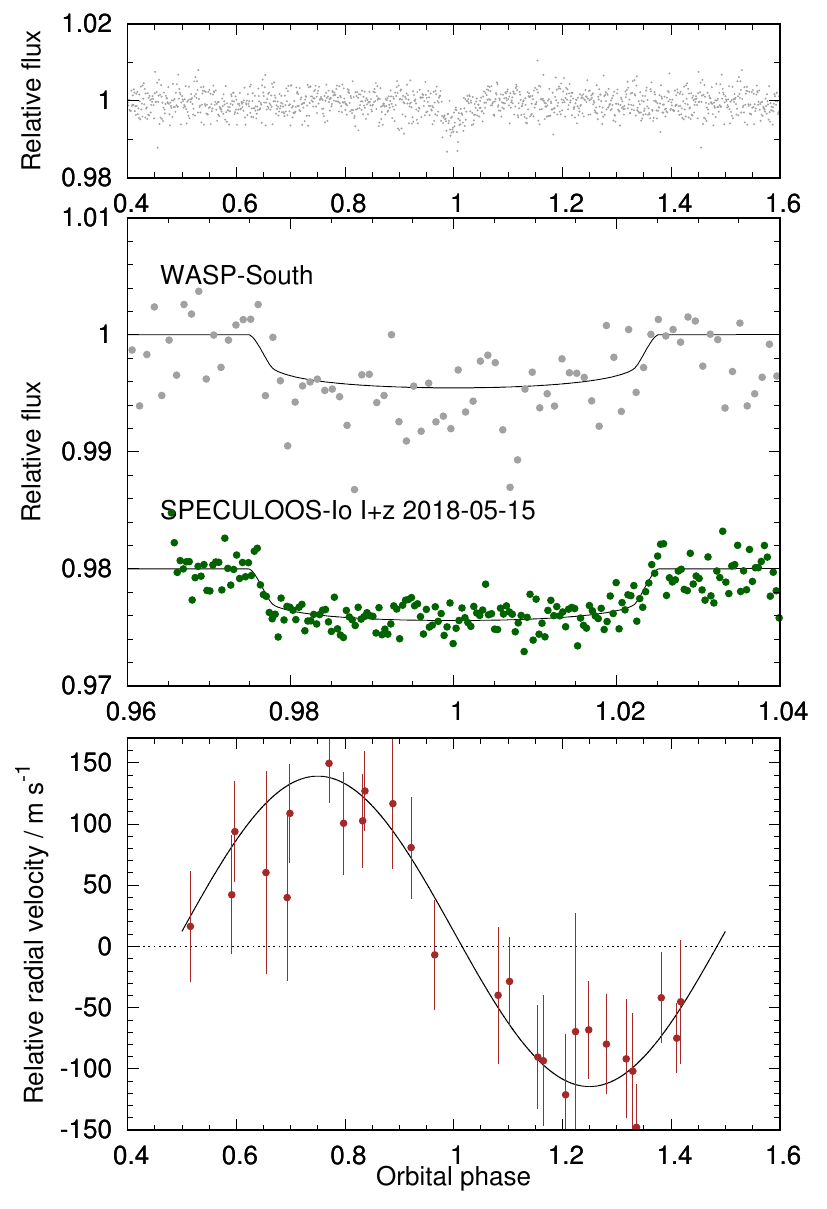}
	\includegraphics[width=0.95\columnwidth]{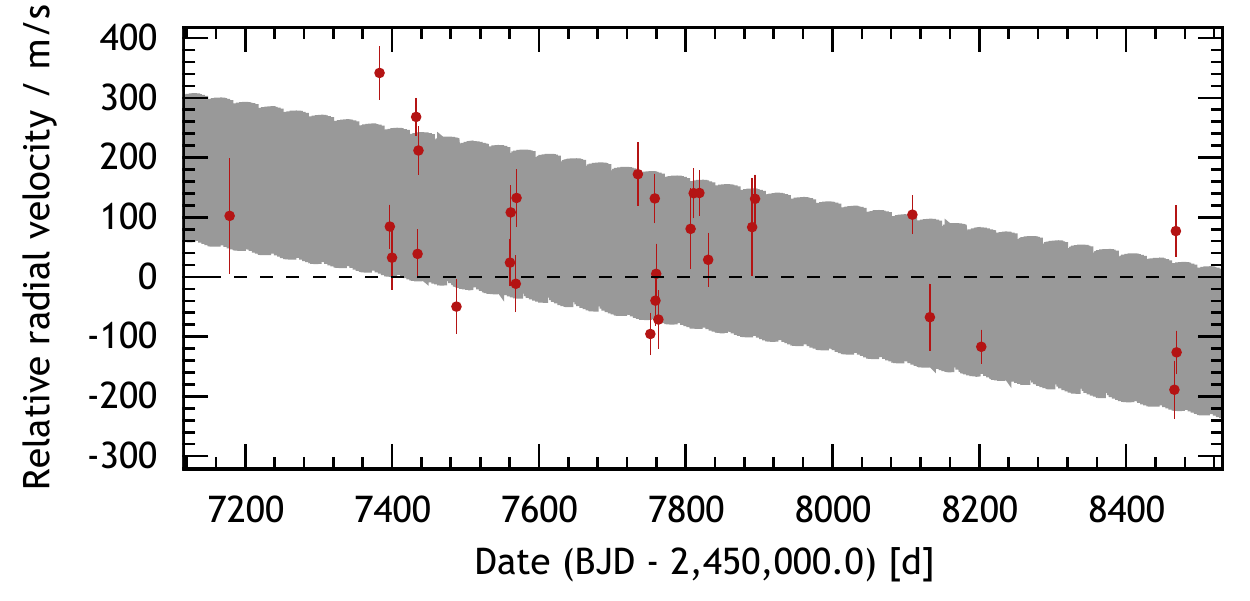}
    \caption{As for Fig. \ref{fig:W169} for the WASP-171 system with the RV-timeseries added in the bottom panel. The best fit Keplerian model is over-plotted with the adopted linear trend. Data from SPECULOOS is shown in green in the  second panel from the top.}
    \label{fig:W171}
\end{figure}

WASP-171 is a V=13.05 G0 star which also appears to be slightly evolved. The transit $\log(g_{s})=4.080^{+0.020}_{-0.049}$ (cgs) is consistent with the spectroscopic value of $4.1 \pm 0.2$. We do find a slight discrepancy between our radius estimate ($1.637^{+0.091}_{-0.046} \mathrm{R_{\odot}}$) and the \gaia\ DR2 value ($2.11^{+0.04}_{-0.2} \mathrm{R_{\odot}}$), though they are consistent to $2\sigma$. The \gaia\ measurements do not seem to be affected by any excess astrometric or phototmetric noise.


WASP-171b is found to have a mass of $1.0841 \pm 0.094~\mathrm{M_{Jup}}$ and radius $0.98^{+0.07}_{-0.04}~\mathrm{R_{Jup}}$, fitting the characteristics of a fairly typical hot Jupiter. The orbital period is 3.82 days, making it the hottest planet presented in this paper with an equilibrium temperature of $T_{eq} = 1640 \pm 40$~K. Figure \ref{fig:W171} shows the \wasp\ discovery light curve with follow-up observations from SPECULOOS-Io and CORALIE. The RVs span a baseline of 3.6 years and show a linear drift of $77 \pm 9 ~\textrm{m s}^{-1}\textrm{yr}^{-1}$, indicating a third body further out in the system. With the data available we can put a minimum mass limit of 10 $\mathrm{M_{Jup}}$ on the outer object, though more observations are needed to constrain whether it is sub-stellar or not.

\subsection{WASP-175b}
\begin{figure}
	\includegraphics[width=\columnwidth]{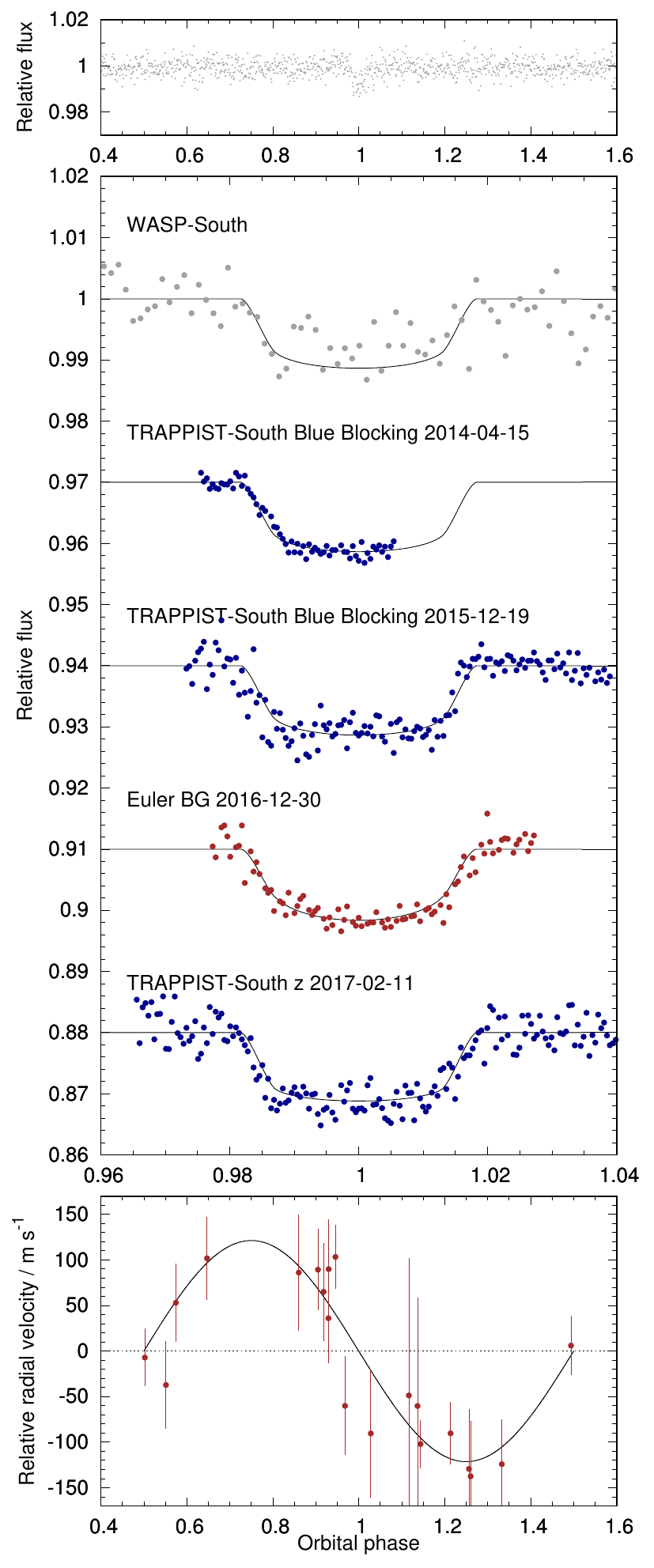}
    \caption{As for Fig. \ref{fig:W169} for the WASP-175 system. Data from EulerCam is shown in red in the middle panel.}
    \label{fig:W175}
\end{figure}

WASP-175 is a V=12.04 F7 star with metallicity [Fe/H]$=0.150 \pm 0.069$. The transit $\log(g_{s})=4.359\pm 0.045$ (cgs) is consistent with the spectroscopic value of $4.3 \pm 0.2$. 

\begin{figure}
\centering
	\includegraphics[width=0.9\columnwidth,trim={11cm 7cm 11cm 7cm},clip]{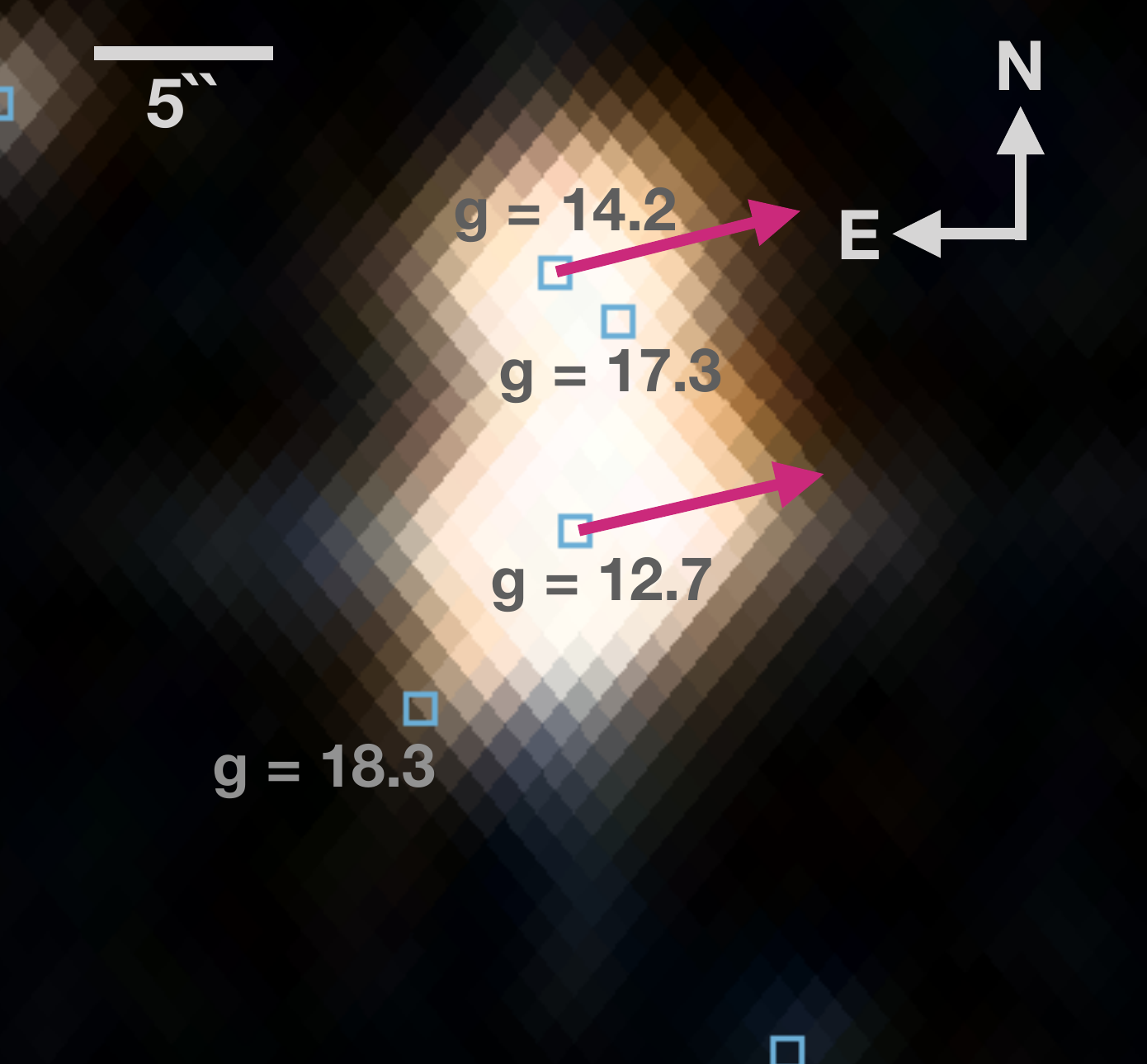}
    \caption{DSS image of WASP-175 (centre) and the nearby companion 7.9\arcsec north. Their common proper motions are indicated as pink arrows. Blue squares are GAIA DR2 sources in the field, with GAIA magnitudes denoted in grey.}
    \label{fig:W175_view}
\end{figure}

Figure \ref{fig:W175} shows the \wasp\ discovery light curve with follow-up observations from TRAPPIST-South, EulerCam and  CORALIE. The \wasp\ light curve is diluted by a star 7.9\arcsec\ away with $\Delta g = 1.5$ \citep{GAIADR2}. The fitted depth of the transit is driven by the follow-up photometry in which the two stars are spatially resolved. The neighbouring star has similar \gaia\ DR2 parallax and is co-moving with WASP-175, indicating that the two stars might be in a wide S-type binary orbit. The projected separation of the two stars is 4600 AU. Fig. \ref{fig:W175_view} shows the two stars with their common proper motion indicated as pink arrows, WASP-175 is the star in the centre of the image. There are two other faint GAIA sources in the field which do not share the same parallax. The companion to the north has GAIA radius $0.88 \pm 0.10~\mathrm{R_{\odot}}$ and effective temperature $5163^{+491}_{-165}~\mathrm{K}$.

WASP-175b has mass $0.99 \pm 0.13~\mathrm{M_{Jup}}$ and radius $1.208 \pm 0.081 ~\mathrm{R_{Jup}}$ and orbits every 3.07 days at a distance of 0.044 AU. Much like the first discovery of a transiting exoplanet \citep[HD 209458b][]{Charbonneau2000} and many more since then, WASP175b fall in the category of hot Jupiters showing anomalous large radii, which cannot be explained by a H-He dominated planet interior \citep{Baraffe2009}. The low density of the planet ($0.56 ^{+0.15}_{-0.11} \mathrm{\rho_{jup}}$) should make WASP-175b a good candidate for atmospheric characterisation. It has an estimated scale height of 620~km, corresponding to a transmission signal of 150~ppm.

\subsection{WASP-182b}
\begin{figure}
	\includegraphics[width=\columnwidth]{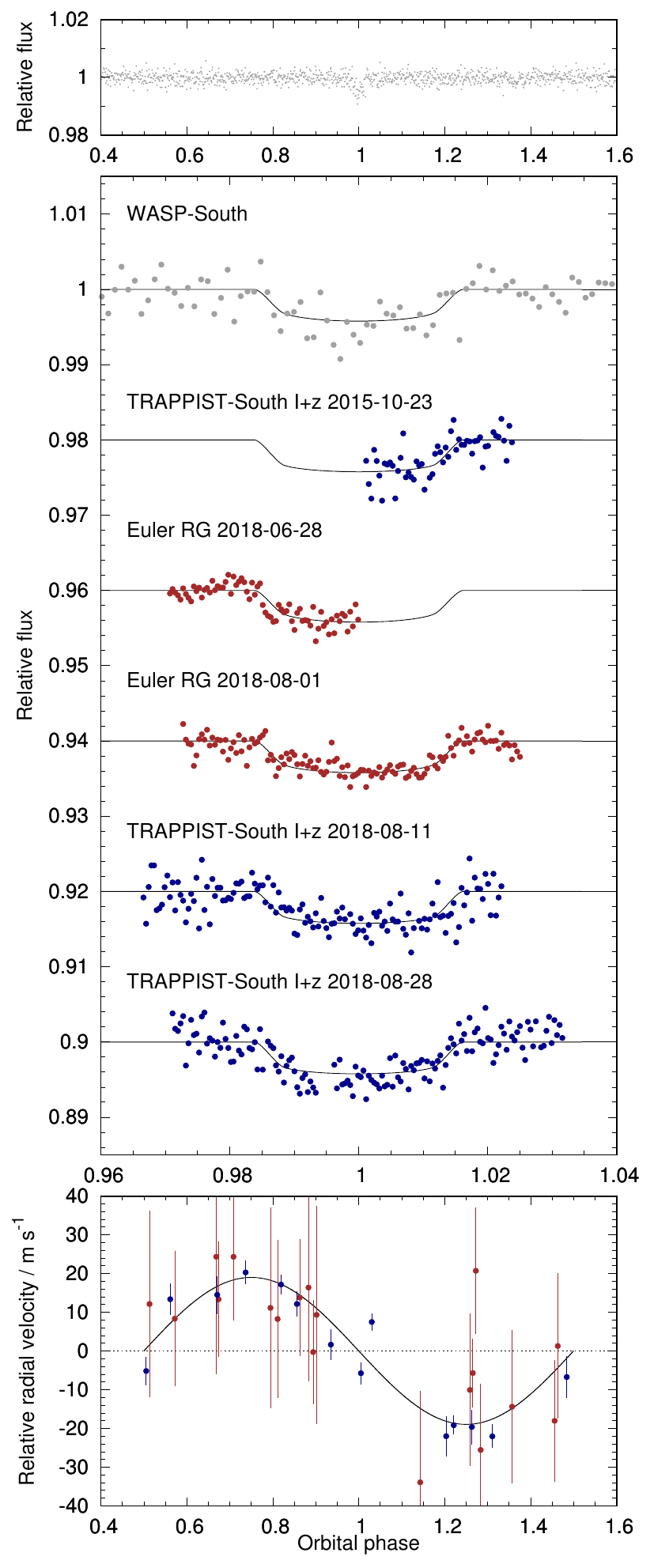}
    \caption{As for Fig. \ref{fig:W175} for the WASP-182 system. Data from HARPS is shown in blue in the bottom panel.}
    \label{fig:W182}
\end{figure}
WASP-182 is a V=11.98 G5 star with a high metallicity, [Fe/H]$=0.27 \pm 0.11$. The stellar density was poorly constrained by the available photometric data, and we thus enforced a prior on the stellar radius from \gaia\ DR2 in the MCMC modelling. The resulting stellar surface gravity  $\log(g_{s})=4.218 \pm 0.033$ (cgs) is consistent with the spectroscopic value of $4.2 \pm 0.2$. 

Figure \ref{fig:W182} shows the \wasp\ discovery light curve with follow-up observations from TRAPPIST-South, EulerCam, CORALIE and HARPS. With a RV semi-amplitude of $19.0 \pm 1.2 ~\mathrm{m s^{-1}}$ a larger telescope was needed to precisely measure the mass of WASP-128b, and we thus obtained data with HARPS as well. The RV scatter around the best fit model is $6.5 ~\mathrm{m s^{-1}}$. One point close to phase=0 (though not in-transit) shows a relatively large offset ($7~\mathrm{m s^{-1}}$) from the joint fit. It does not appear to be affected by stellar activity or other systematic effects, so we have included it in the analysis for completion. 
Using the \wasp\ photometry we constrain the stellar rotation period to $30 \pm 2~{\rm days}$, which is consistent with a G-star on the main sequence \citep{McQuillan2014}. The RVs show no variability, as a sign of stellar activity, at that period.

WASP-182b is found to have a mass of $0.148 \pm 0.011~\mathrm{M_{Jup}}$ and radius $0.850 \pm 0.030~\mathrm{R_{Jup}}$, making it a low density planet. The estimated scale height is 1930~km, corresponding to 264~ppm. With a period of 3.38 days WASP-182b sits right between the lower and upper edges of the sub-Jovian desert in both the mass- and radius plane, as seen in Fig.~\ref{fig:MR}. This makes it an even more compelling target for in-depth atmospheric characterisation, studying possible atmospheric escape close to the evaporation desert \citep{Owen2018,Ehrenreich2015,Bourrier2018}.

\section{Discussion \& Conclusion}
\begin{figure}
	\includegraphics[width=\columnwidth,trim={0.2cm 0.4cm 0cm 0cm},clip]{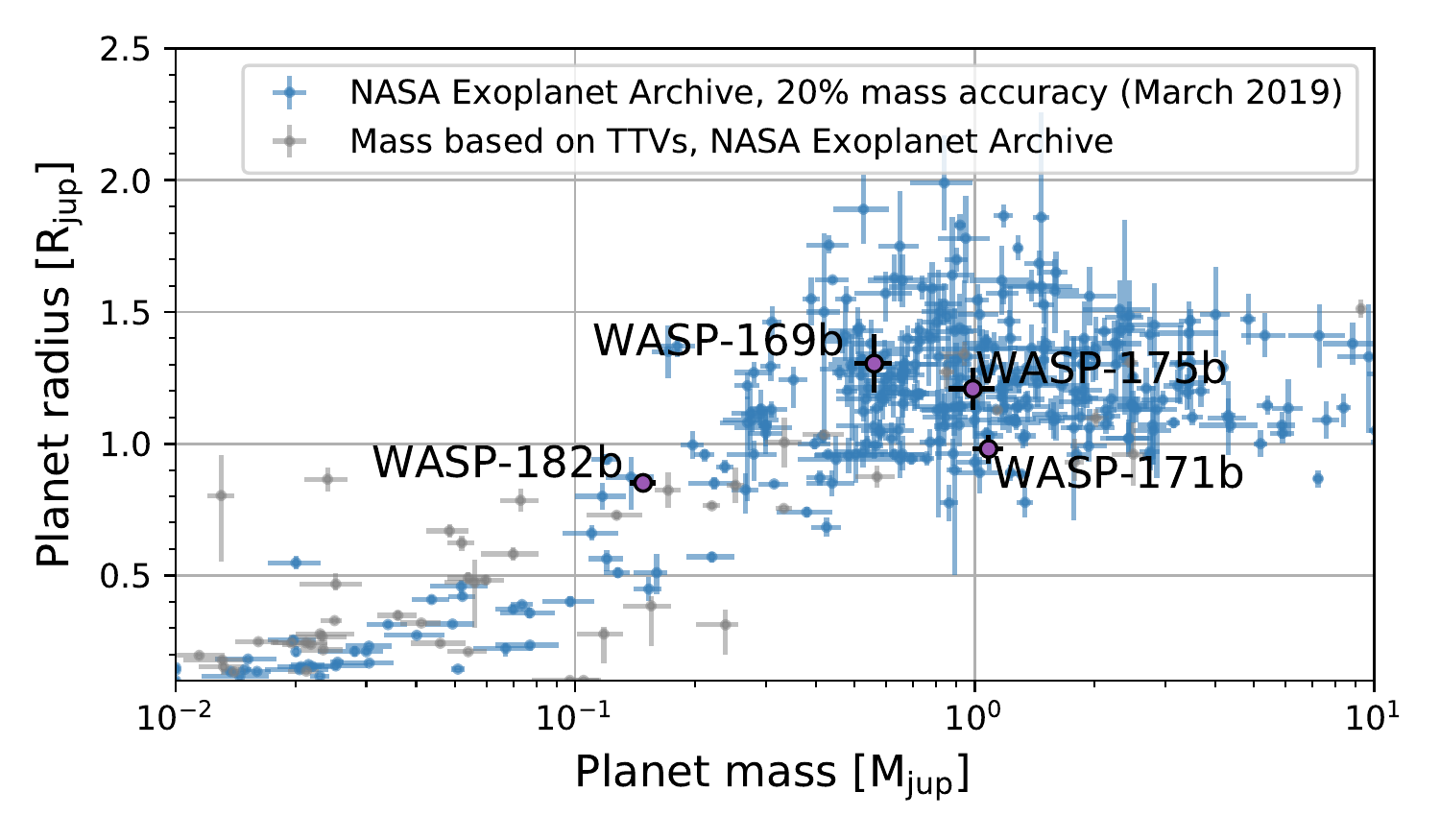}
\caption{Masses and radii of the four planets presented in this paper, WASP-169b, WASP-171b, WASP-175b and WASP-182b, along with the known exoplanet population. Only planets with masses determined to 20\% or better are included, and mass-estimates based on TTVs are in grey.}
    \label{fig:MR}
\end{figure}
We have presented the discovery and mass determination of four new Jovian planets from the \wasp\ survey. Fig. \ref{fig:MR} presents these planets along with the mass and radii of the known exoplanet population, as per March 2019. Only planets with masses determined to a fractional accuracy of 20\% or better are included, and mass-estimates based on transit-timing variations (TTVs) are distinguished in grey. 

WASP-169b, WASP-171b and WASP-175b fall within the category of hot Jupiters, with WASP-169b and WASP-175b being inflated. Having precise parameters for inflated Jupiters across a variety of stellar hosts and evolutionary stages will help to solve the conundrum of the hot Jupiter radius-anomaly. 

WASP-182b is a bloated sub-Saturn mass planet, occupying a poorly populated parameter-space, corresponding to the transition between Neptune-like ice-giants and Saturn-like gas-giants, at $0.05 - 0.3~\mathrm{M_{Jup}}$. Less than 30 planets in this range have masses determined to 20\% fractional accuracy or better. 

\begin{figure}
    \includegraphics[width=\columnwidth,trim={0.2cm 0.4cm 0cm 0cm},clip]{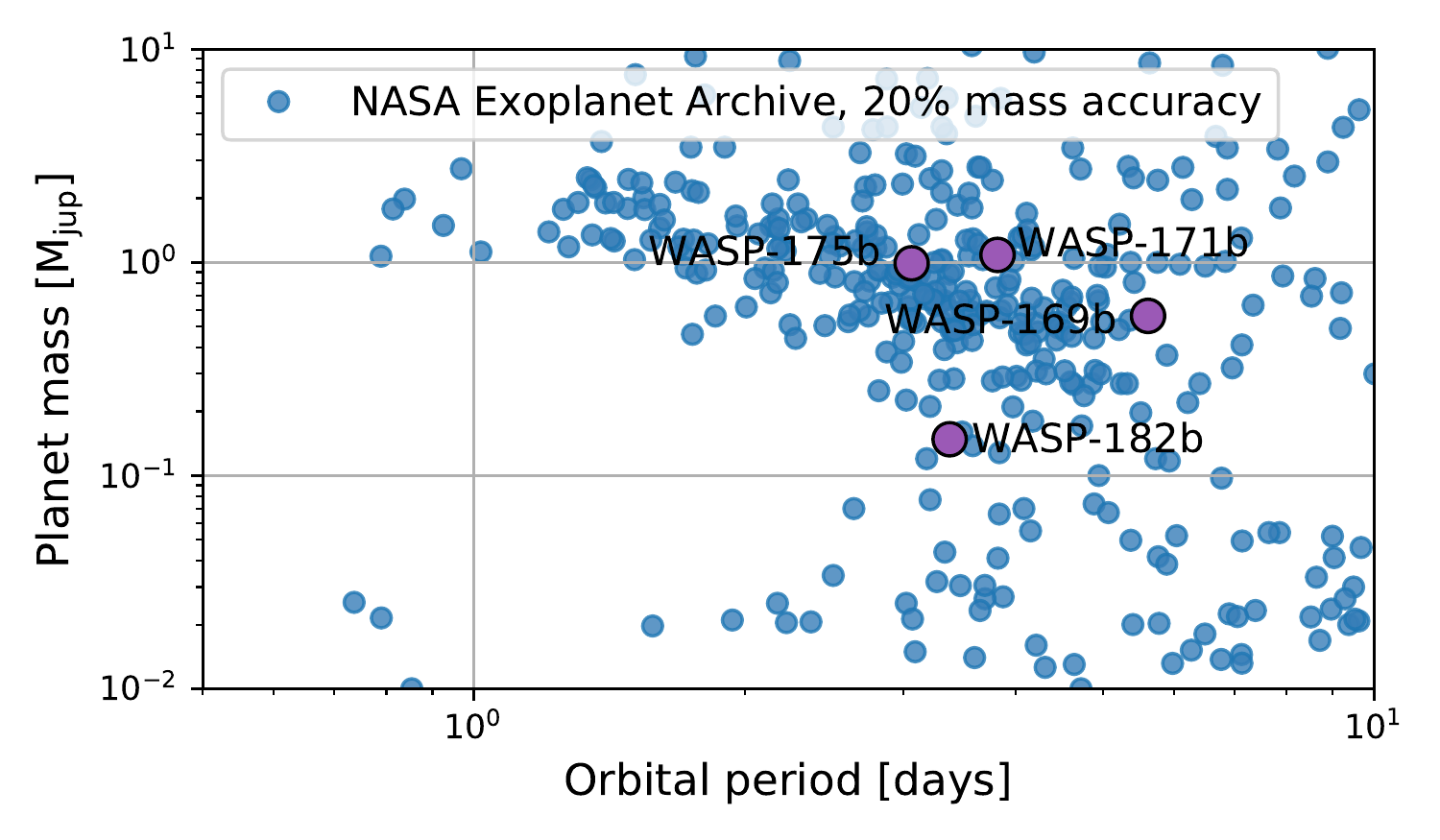}
    \includegraphics[width=\columnwidth,trim={0.2cm 0.4cm 0cm 0cm},clip]{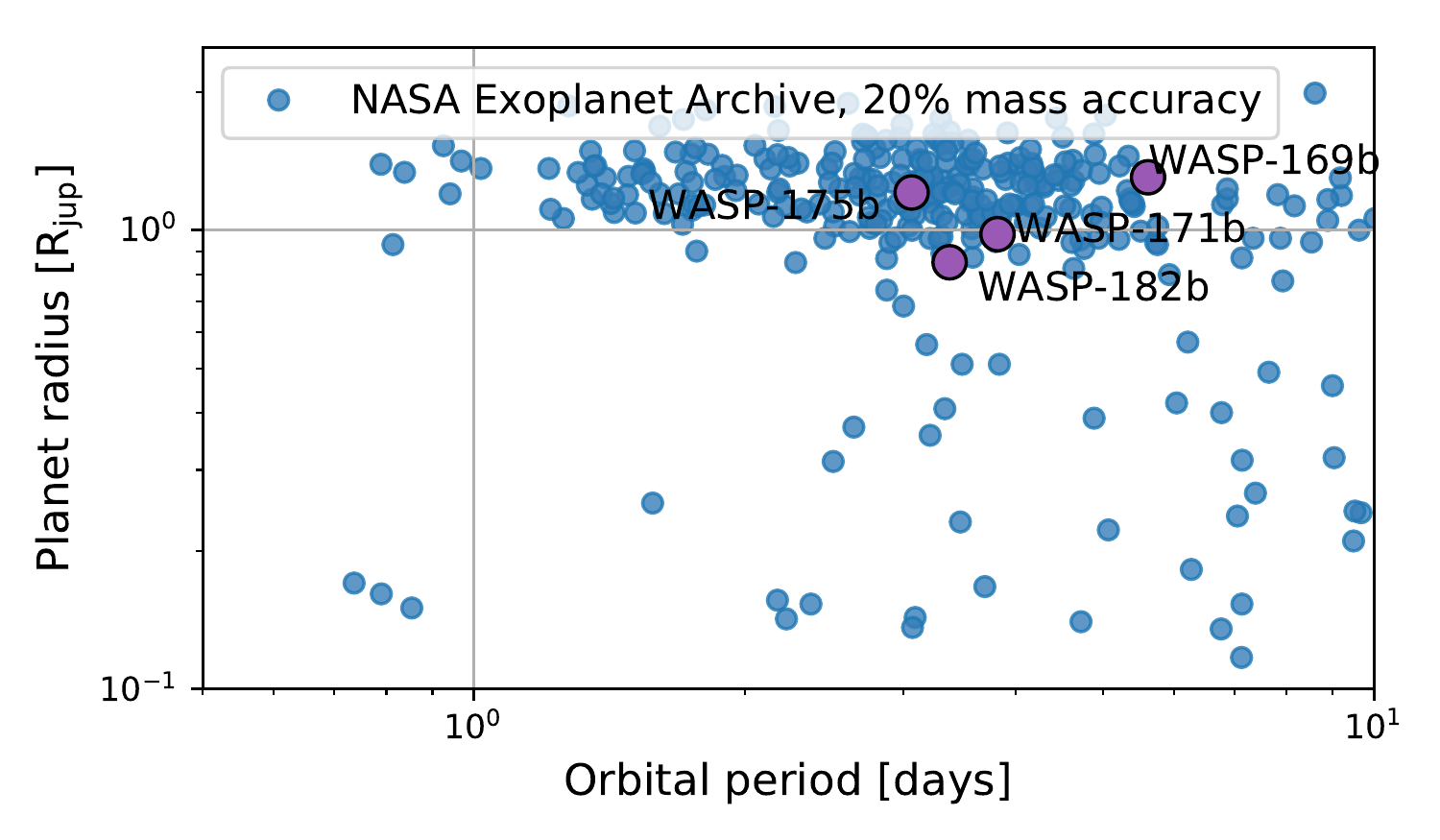}
\caption{ The sub-Jovian deserts illustrated through the know exoplanet population with masses determined to 20\% or better, as in Fig. \ref{fig:MR}.  \textit{Top panel} show the mass-period plane, whereas the \textit{bottom panel} shows the same dearth of sub-jovian planets with short periods in the radius- period plane. WASP-182b sits at the apex of the sub-Jovian desert in both the mass- and radius planes.}
    \label{fig:MRP}
\end{figure}
 Furthermore, WASP-182b sits right in the apex of the sub-Jovian desert, as defined by \cite{Mazeh2016,Szabo2011}, see Fig. \ref{fig:MRP}. The proposed mechanisms behind this dearth of sub-Saturn planets with short periods are numerous, but can generally be classified as being related to disk-material available during planet formation or photo evaporation for the small planets. For the larger ones, framing the top of the desert, migration of massive planet from further out in the system could allow the most massive objects to keep their atmospheric volatile layer as they approach the host star \citep{2014ApJ...792....1L, 2015IJAsB..14..201M}. Whereas less massive planets will loose their outer layer and perhaps end up as a naked core in the bottom of the desert \citep{2018MNRAS.479.5012O}. Finding planets such as WASP-182b that sits between the two edges will help identify the most important physical processes behind the desert.

\section*{Acknowledgements}

We thank the Swiss National Science Foundation (SNSF)
and the Geneva University for their continuous support to our planet search programs. This work has been in particular carried out in the frame of the National Centre for Competence in Research `PlanetS' supported by the Swiss National Science Foundation (SNSF). 

This publication makes use of The Data \& Analysis Center for Exoplanets (DACE), which is a facility based at the University of Geneva (CH) dedicated to extrasolar planets data visualisation, exchange and analysis. DACE is a platform of the Swiss National Centre of Competence in Research (NCCR) PlanetS, federating the Swiss expertise in Exoplanet research. The DACE platform is available at \url{https://dace.unige.ch}.

WASP-South is hosted by the South African Astronomical Observatory and we are grateful for their ongoing support and assistance.
Funding for WASP comes from consortium universities and from the UK's Science and Technology Facilities Council. 
TRAPPIST is funded by the Belgian Fund for Scientific Research (Fond National de la Recherche Scientifique, FNRS) under the grant FRFC 2.5.594.09.F, with the participation of the Swiss National Science Fundation (SNF). MG and EJ are F.R.S.-FNRS Senior Research Associates.

The research leading to these results has received funding from the European Research Council under the FP/2007-2013 ERC Grant Agreement 336480, from the ARC grant for Concerted Research Actions  financed  by  the  Wallonia-Brussels  Federation,  from the Balzan Foundation, and a grant from the Erasmus+ International Credit Mobility programme (K. Barkaoui).




\bibliographystyle{mnras}
\bibliography{wasp_nielsen} 








\bsp	
\label{lastpage}

\end{document}